\documentclass{article}
\usepackage{graphics}
\usepackage{newfloat}
\usepackage{listings}
\usepackage{algorithm}
\usepackage{algorithmic}
\usepackage{array}
\usepackage{multirow}
\usepackage{multicol}
\usepackage{latexsym}
\usepackage{stfloats}
\usepackage{paper,amsmath,graphicx}

\title{DSNet: a simple yet efficient network with dual-stream attention for lesion segmentation}
\name{Yunxiao Liu}
\address{Tsinghua University\\
	Shenzhen International Graduate School\\
	liuyx\_21@163.com}
\begin{document}
\maketitle
\begin{abstract}
Lesion segmentation requires both speed and accuracy. In this paper, we propose a simple yet efficient network DSNet, which consists of a encoder based on Transformer and a convolutional neural network(CNN)-based distinct pyramid decoder containing three dual-stream attention (DSA) modules. Specifically, the DSA module fuses features from two adjacent levels through the false positive stream attention (FPSA) branch and the false negative stream attention (FNSA) branch to obtain features with diversified contextual information. We compare our method with various state-of-the-art (SOTA) lesion segmentation methods with several public datasets, including CVC-ClinicDB, Kvasir-SEG, and ISIC-2018 Task 1. The experimental results show that our method achieves SOTA performance in terms of mean Dice coefficient (mDice) and mean Intersection over Union (mIoU) with low model complexity and memory consumption.
\end{abstract}
\begin{keywords}
Lesion segmentation, Attention, Transformer, Convolutional neural network
\end{keywords}
\section{Introduction}
\label{sec:intro}
Lesion segmentation is one of the primary objectives of medical imaging, since the size and location of lesions are commonly directly associated with the diagnosis, treatment and prognosis of patients\cite{fully}. Traditionally, the location and size of lesions are manually determined by experts, which means that the results of diagnosis usually depend on the experience of the experts and can be time-consuming, subjective and error-prone\cite{contour}. With the rapid development of computer vision algorithms, researchers have begun to apply these algorithms in lesion segmentation to assist diagnosis\cite{Statistical}. \\
Classical lesion segmentation algorithms are based on threshold, boundary, region growth, fuzzy set theory, etc. Nevertheless, these algorithms generally rely heavily on prior knowledge, require manual design of features, and have limited performance on complex datasets\cite{Statistical}. Therefore, it is essential to construct a universal, efficient and robust lesion segmentation algorithm.
\begin{figure*}[ht]
\centering
\includegraphics[width=0.95\textwidth]{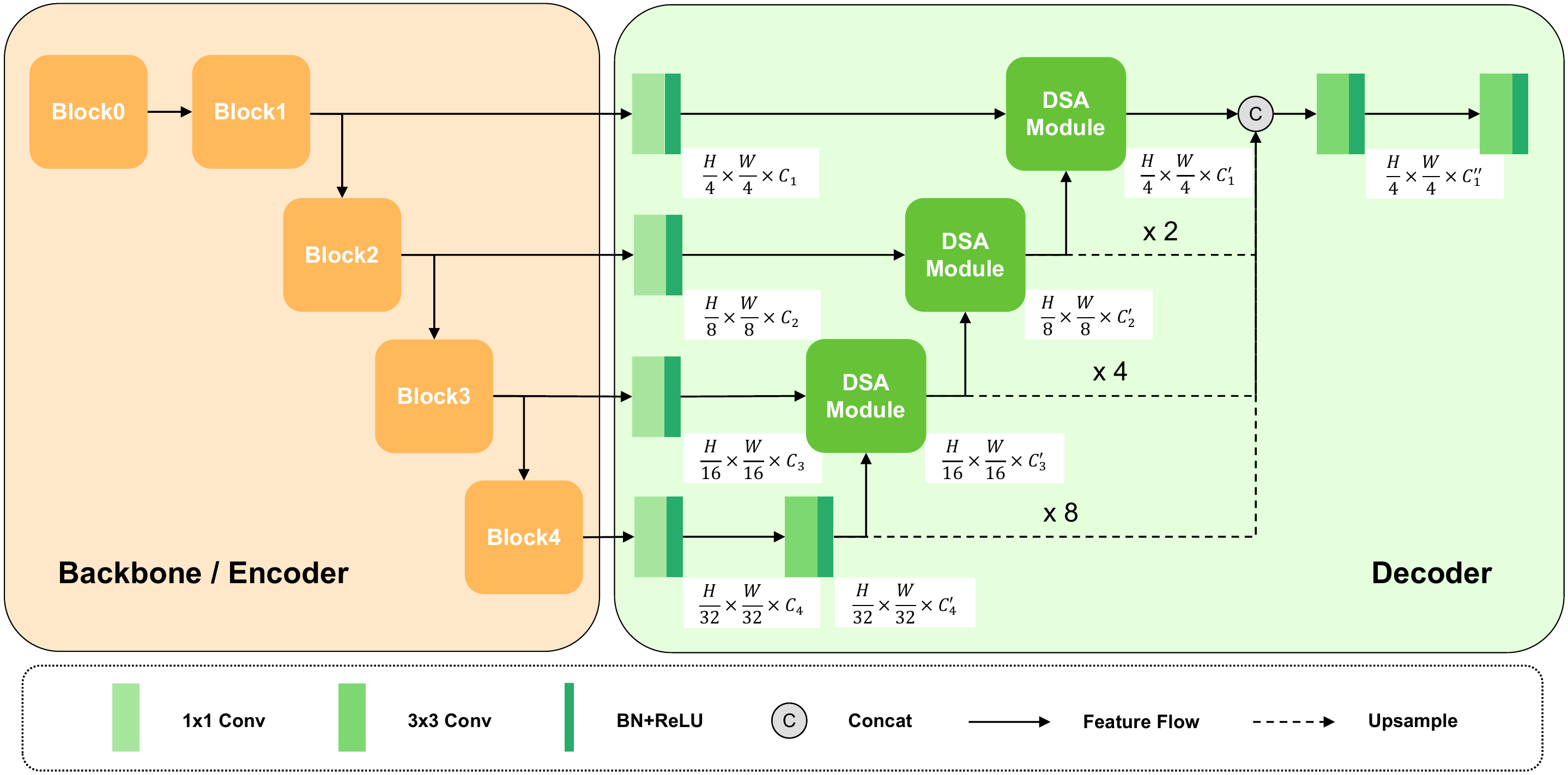} 
\caption{An overview of DSNet, which contains a pre-trained backbone encoder and a pyramid decoder with DSA modules.}
\label{fig1: Overview of our model}
\end{figure*}\\
After FCN\cite{FCN}, especially U-Net\cite{unet}, was proposed, CNN-based encoder-decoder architectures have achieved significantly superior performance than classical algorithms in lesion segmentation\cite{fully}. The success of these models depends largely on their employment of skip connections in decoders to fuse fine-grained semantic features with coarse-grained semantic features to generate the final output\cite{multi}. Therefore, many researchers have focused on increasing various skip connections to improve performance, such as UNet++\cite{unet++}, UNet3+\cite{unet3p}. Although these improvements are effective, abundant skip connections generally lead to an undesirable increase in the computational complexity of models. Our method also employs the encoder-decoder architecture, where we utilize as few skip connections as possible in the decoder to keep it simple and efficient.\\
Some researchers recognized that the main drawback of CNN was the inability to capture long range dependencies, such as the non-local correlation of objects and the extraction of contextual information\cite{ViTreview}. This triggered attempts to incorporate attention mechanisms into CNN architecture to improve performance, such as integrating channel-wise attention\cite{woo2018cbam}\cite{hu2018squeeze} and spatial attention\cite{cao2019gcnet}\cite{huang2019ccnet}. CNN architecture with various attention mechanisms, such as Attention-UNet\cite{attentionunet}, achieved SOTA performance in lesion segmentation for a while. Recently, Dosovitskiy et al. proposed Vision Transformer(ViT) based on the Transformer architecture used in natural language processing tasks\cite{ViT}. ViT and its variants\cite{PVT}\cite{swin}\cite{scaling} can capture long range dependencies more robustly than CNN, which has led researchers to start using them as the encoder of lesion segmentation models to extract features. Our method also utilizes a variant of ViT, Mix Transformer (MiT)\cite{segformer}, as the encoder to capture features with long range dependencies. Meanwhile, we propose a novel attention module, DSA module, in decoder part whose component and function are typically different from those of previous attention modules.\\
The main contributions of our work can be summarized as follows: 
\begin{itemize}
\item[$\bullet$] We propose a novel attention module, DSA module, whose FPSA branch and FNSA branch mimics the human process of carefully analyzed interference elimination to obtain features with diversified contextual information.
\item[$\bullet$] We implement a simple yet efficient network, DSNet, with the MiT backbone encoder and the pyramid decoder based on DSA module.
\item[$\bullet$] We conduct experimental analysis in three public lesion segmentation datasets. Evaluation results demonstrate that our DSNet achieves SOTA performance in terms of mDice, mIoU, the number of parameters and floating point of operations, which can be a new SOTA method for lesion segmentation task.
\end{itemize}
\section{Method}
\label{sec:majhead}
As illustrated in Fig. 1, DSNet employs an encoder-decoder architecture, where the encoder is a pre-trained MiT backbone and the decoder is a simple pyramid CNN architecture containing three DSA modules. In the following subsections, we will describe the components of DSNet in detail.
\subsection{Encoder}
\label{ssec:subhead}
In the segmentation task, a powerful encoder is necessary, hence our encoder is MiT, a Transformer backbone with excellent performance proposed in SegFormer\cite{segformer}. MiT improves the patch embedding, named Block 0 in Fig. 1, of ViT such that there is overlap among adjacent patches to ensure local continuity. Meanwhile, MiT removes the position embedding and introduces deepwise convolution into the feed forward network to transmit the position information, which improves the resolution of the input image of MiT variable. MiT has four stages, named Block 1 to 4 in Fig. 1, each of which furnishes features of different resolution and granularity for downstream tasks. According to the mechanism of transfer learning, we use MiT weights pre-trained on the ImageNet database\cite{imagenet} to enhance the performance of the model on data-hunger lesion segmentation datasets.
\begin{figure}[htb]
  \centering
  \centerline{\includegraphics[width=8.5cm]{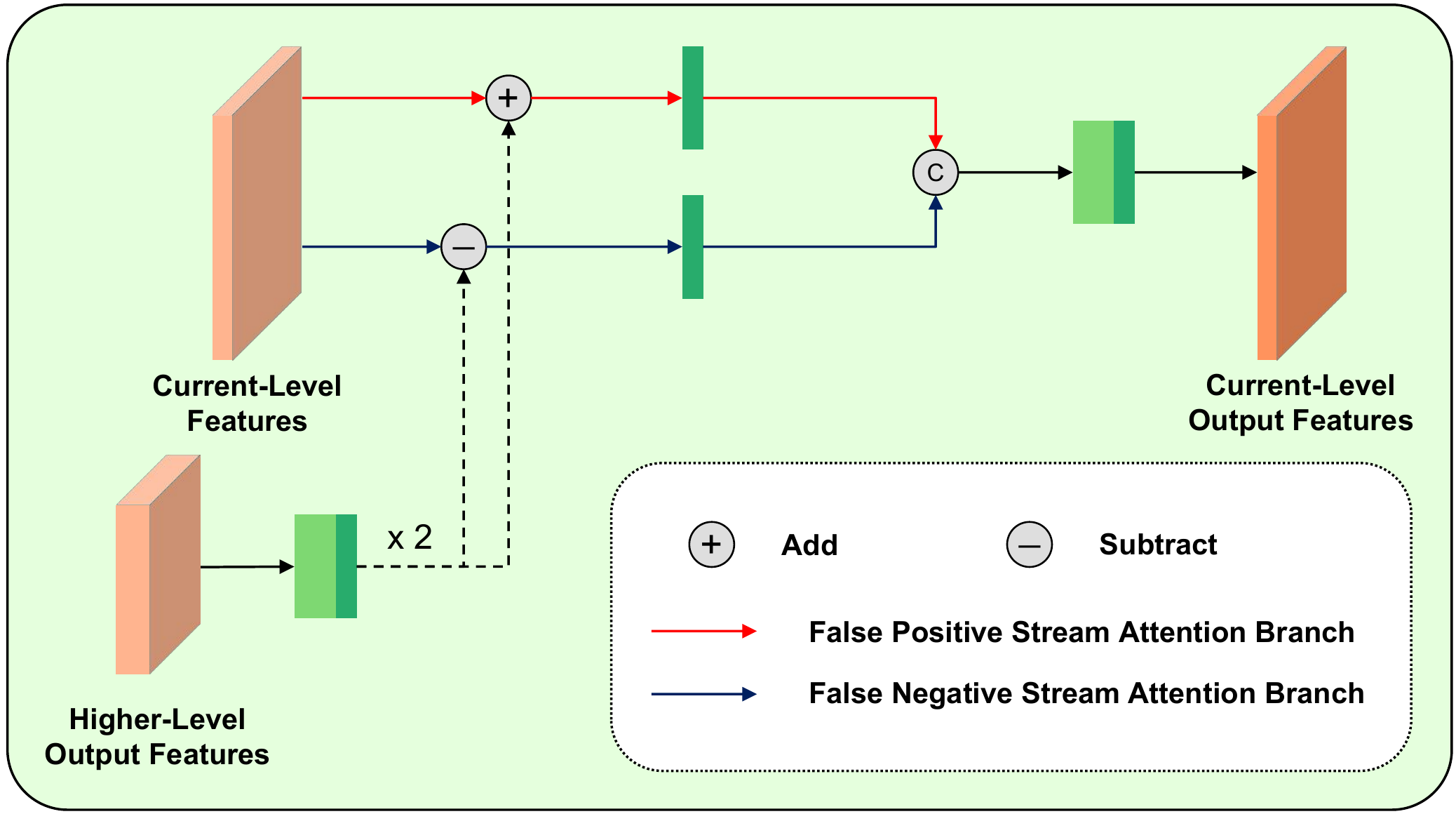}}
\caption{Details of our DSA Module.}
\label{fig2: Overview of our DS Module}
\end{figure}
\subsection{Decoder}
In both SegFormer\cite{segformer} and ESFPNet\cite{ESFP}, the decoders use Multi-Layer Perceptrons (MLP) to fuse features to reduce the number of parameters. Although the MLP-based decoders are effective, they performs features fusion considering less local information. To adequately exploit the local information while maintaining low computational complexity, our decoder is based on a simple CNN architecture. First, the output features of different stages of the encoder will be delivered to $1\times{1}$ convolutions to reduce the number of channels. All convolutions in our method are followed by BatchNorm and ReLU activation (BR) operations. Then, the highest-level features will go through a $3\times{3}$ convolution to further extract the local information, and the features of other layers will be sent to the DSA modules together with the adjacent higher-layer output features to obtain current-level output features. Finally, We concatenate these output features containing diversified contextual information, further fuse them through two $3\times3$ convolutions to get the final output.\\
 Our novel proposed DSA module, shown in Figure 2, is inspired by PFNet\cite{pfnet} and consists of a FPSA branch and a FNSA branch, which mimics the human process of eliminating interference by carefully analyzing false positive and false negative errors. However, compared to PFNet, DSA module is much simpler and lighter. For each attention branch in DSA module, the features with false positive or false negative attention are obtained by adding or subtracting the higher-level output features processed by a $3\times3$ convolution and then following a BR. Combined with our proposed DSA module, the decoder presents a stronger ability to focus on diversified contextual information to guide the final segmentation decision.
\label{sssec:subsubhead}
\section{EXPERIMENTS and Results}
\label{sec:print}
\subsection{Datasets}
To demonstrate the effectiveness of DSNet, we evaluate it on three public lesion segmentation datasets, CVC-ClinicDB\cite{CVC}, Kvasir-SEG\cite{SEG} and ISIC 2018 Task 1\cite{skin_1, skin_2}. CVC-ClinicDB and Kvasir-SEG are commonly-used datasets for the polyp segmentation task, and ISIC 2018 is a large-scale dataset of skin dermoscopy images, of which Task 1 is a challenge on melanoma segmentation. Following ESFPNet\cite{ESFP}, we resize the images to $352\times{352}$ for CVC-ClinicDB and Kvasir-SEG. For ISIC 2018 Task 1, the input images are resized to $512\times{512}$. The details about data split are presented in Table 1.
\begin{table}[htbp] 
\centering 
\caption{Details of the lesion segmentation datasets used in our experiments.}
\resizebox{\columnwidth}{!}
{\begin{tabular}{m{2.3cm}<{\centering} m{0.75cm}<{\centering} m{1.65cm}<{\centering} m{0.5cm}<{\centering} m{0.5cm}<{\centering} m{0.5cm}<{\centering}}
\hline
Dataset & Images & Image size & Train & Valid & Test  \\ \hline
CVC-ClinicDB & 612 & $352\times{352}$ & 490 & 61 & 61  \\ \hline
Kvasir-SEG & 1000 &	$352\times{352}$ & 800 & 100 & 100 \\ \hline
ISIC-2018 & 2594 & $512\times{512}$ & 1868 & 467 & 259 \\ \hline
\end{tabular}}
\end{table}
\subsection{Implementation Details}
We propose four DSNet models with successively increasing scales, DSNet-T (Tiny DSNet), DSNet-S (Small DSNet), DSNet-B (Base DSNet), and DSNet-L (Large DSNet), which use MiT backbone B0 to B3 respectively. For $C_{1}$, $C_{2}$, $C_{3}$, $C_{4}$ in Fig 1, we set them $64$, $128$, $128$, $256$, and we set $C_{1}^{'}$, $C_{2}^{'}$, $C_{3}^{'}$, $C_{4}^{'}$, $C_{1}^{''}$ to $32$, $64$, $96$, $96$, $256$. We implement DSNet using the PyTorch framework on two NVIDIA GeForce RTX 2080Ti GPUs (with 11 GB memory). During training, we employ random brightness, flipping, and rotation changing as data augmentation operations. The loss function we use is the combination of the binary cross-entropy (BCE) loss and the Dice loss. Then we train DSNet on each dataset for 200 epochs with a batch size of 8 and the AdamW optimizer\cite{AdamW}. The initial learning rate is 1e-4. and we apply ReduceLROnPlateau to optimise the learning rate. We reduce the learning rate by a factor of 2 when the Dice coefficient on the validation set do not improve over 10 epochs until reaching a minimum of 1e-6, and save the model if the performance improves.
\begin{table}[htbp]
\centering
\caption{Quantitative results on CVC-ClinicDB.}
\resizebox{\columnwidth}{!}
{\begin{tabular}{m{3.0cm}<{\centering} m{0.9cm}<{\centering} m{0.9cm}<{\centering} m{1.2cm}<{\centering} m{1.6cm}<{\centering}}
\hline
Network & mDice$\uparrow$ & mIoU$\uparrow$ & Para(M)$\downarrow$ & FLOPs(G)$\downarrow$  \\ \hline
UNet++\cite{unet++} & 0.794 & 0.755 & 9.2 & 65.9  \\ \hline 
DCSAU-Net\cite{dcsau} & 0.916 & 0.861 & \pmb{2.6} & \pmb{12.9}  \\ \hline
CaraNet\cite{caranet} & 0.936 & 0.887 & 44.6 & 21.7  \\ \hline 
ESFPNet-L\cite{ESFP} & 0.949 & 0.907 & 61.7 & 23.9  \\ \hline
FCBForemer-L\cite{FCBFormer} & 0.947 & 0.902 & 52.9 & 73.4  \\ \hline
DSNet-L(Ours) & \pmb{0.950} & \pmb{0.908} & 45.6 & 23.5 \\ \hline
\end{tabular}}
\end{table}
\subsection{Results}
In this section, we present quantitative results on three public datasets and compare DSNet with other SOTA methods with respect to mDice, mIoU, the number of parameters (Para), and floating point of operations (FLOPs). We utilize one input image to calculate FLOPs in our experiment.
\begin{table}[htbp]
\centering
\caption{Quantitative results on Kvasir-SEG.}
\resizebox{\columnwidth}{!}
{\begin{tabular}{m{3.0cm}<{\centering} m{0.9cm}<{\centering} m{0.9cm}<{\centering} m{1.2cm}<{\centering} m{1.6cm}<{\centering}}
\hline
Network & mDice$\uparrow$ & mIoU$\uparrow$ & Para(M)$\downarrow$ & FLOPs(G)$\downarrow$  \\ \hline
PraNet\cite{pranet} & 0.898 & 0.849 & \pmb{30.5} & \pmb{13.1}  \\ \hline
Deeplabv3+\cite{deeplab} & 0.897 & 0.858 & 39.0 & 22.7  \\ \hline
GMSRF-Net\cite{gmsrf} & 0.926 & 0.884 & 47.0 & 41.5  \\ \hline
SSFormer-L\cite{ssformer} & 0.936 & 0.891 & 66.2 & 34.6  \\ \hline
ESFPNet-L\cite{ESFP} & 0.931 & 0.887 & 61.7 & 23.9  \\ \hline
FCBForemer-L\cite{FCBFormer} & \pmb{0.939} & 0.890 & 52.9 & 73.4  \\ \hline
DSNet-L(Ours) & \pmb{0.939} & \pmb{0.893} & 45.6 & 23.5 \\ \hline
\end{tabular}}
\end{table}\\
The quantitative results on CVC-ClinicDB dataset are shown in Table 2. The performance of our method on mDice and mIoU is essentially the same as that of the most advanced ESFPNet, yet our method outperforms it in terms of Para and FLOPs, which is a testament to the effectiveness of our decoder since DSNet-L employ MiT B3 as encoder while ESFPNet-L utilizes MiT B4. Although our method has additional Para and larger FLOPs compared to DCSAU-Net and U-net++, our method has a clear advantage in mDice and mIoU.
\begin{table}[htbp]
\centering
\caption{Quantitative results on ISIC-2018 Task 1.}
\resizebox{\columnwidth}{!}
{\begin{tabular}{m{3.0cm}<{\centering} m{0.9cm}<{\centering} m{0.9cm}<{\centering} m{1.2cm}<{\centering} m{1.6cm}<{\centering}}
\hline
Network & mDice$\uparrow$ & mIoU$\uparrow$ & Para(M)$\downarrow$ & FLOPs(G)$\downarrow$  \\ \hline 
U-Net\cite{unet} & 0.890 & 0.802 & 7.8 & 54.9  \\ \hline
UNet3+\cite{unet3p} & 0.899 & 0.816 & 27.0 & 799.0  \\ \hline
Attention-UNet\cite{attentionunet} & 0.897 & 0.814 & 34.9 & 266.3  \\ \hline
DCSAU-Net\cite{dcsau} & 0.914 & 0.841 & \pmb{2.6} & \pmb{27.2}  \\ \hline
DSNet-L(Ours) & \pmb{0.919} & \pmb{0.860} & 45.6 & 45.4 \\\hline
\end{tabular}}
\end{table}\\
As shown in Table 3, our method achieves the best performance in terms of mDice and mIoU on Kavsir-SEG dataset, and our method is superior to SSFormer, ESFPNet and GMSRF-Net on all four metrics.\\
From the results shown in Table 4, for ISIC-2018 Task 1, our method also achieves the best results on mDice and mIoU, which demonstrates the generality of our method. Besides, compared with U-Net, UNet3+, and Attention-UNet, although they have fewer Para, they suffer from larger FLOPs due to their complex decoders with plenty of convolutions and skip connections to compensate for the inability of their encoders to capture long range dependencies in features extraction. The MiT encoder can efficiently extract features with rich global information, and the simplified pyramid decoder with DSA modules can fully integrate contextual information, which makes our method perform excellently on FLOPs.
\begin{table*}[hb!]
\centering
\caption{Ablation results on Kvasir-SEG. "B" denotes DSNet without attention branch in DSA modules, "+FP" denotes DSNet with FPSA branch, and "+FN" denotes DSNet with FNSA branch.}
{\begin{tabular}{m{2.0cm}<{\centering} m{1.0cm}<{\centering} m{1.0cm}<{\centering} m{1.0cm}<{\centering} m{1.0cm}<{\centering} m{1.0cm}<{\centering} m{1.0cm}<{\centering} m{1.0cm}<{\centering} m{1.0cm}<{\centering} m{1.5cm}<{\centering} m{1.5cm}<{\centering}}
\hline
\multirow{2}{*}{Network} & \multicolumn{2}{c}{B} & \multicolumn{2}{c}{+FP}  & \multicolumn{2}{c}{+FN} & \multicolumn{4}{c}{+FP+FN} \\ \cline{2-11}
 & mDice$\uparrow$ & mIoU$\uparrow$ & mDice$\uparrow$ & mIoU$\uparrow$ & mDice$\uparrow$ & mIoU$\uparrow$ & mDice$\uparrow$ & mIoU$\uparrow$ & Para(M)$\downarrow$ & FLOPs(G)$\downarrow$ \\ \hline
DSNet-L & 0.924 & 0.872 & 0.931 & 0.887 & 0.926 & 0.877 & \pmb{0.939} & \pmb{0.893} & 45.6 & 23.5\\ \hline
DSNet-B & 0.921 & 0.868 & \pmb{0.929} & 0.879 & 0.921 & 0.871 & \pmb{0.929} & \pmb{0.880} & 25.8 & 14.5\\ \hline
DSNet-S & 0.912 & 0.859 & 0.912 & 0.859 & 0.917 & 0.865 & \pmb{0.921} & \pmb{0.870} & 14.7 & 10.8 \\ \hline
DSNet-T & 0.911 & 0.856 & 0.912 & 0.858 & 0.907 & 0.855 & \pmb{0.915} & \pmb{0.863} & 4.8 & 7.8\\ \hline
\end{tabular}}
\end{table*}
\subsection{Ablation Studies}
To further demonstrate the validity of our method, we conduct ablation studies with Kvasir-SEG dataset. We replace DSA modules with $3\times3$ convolutions as the baseline, and conduct ablation on attention branches of DSA modules. From Table 5, it can be seen that integrating either the FPSA branch or the FNSA branch to the baseline improves performance and DSNet achieves the best performance on all scales when both the FPSA branch and the FPSA branch are used, which definitely illustrates the effectiveness of our design for DSA module.  Moreover, compared with other methods in Table 3, DSNet-T, DSNet-S and DSNet-B also achieve competitive performance, which provides more options for DSNet applications.
\section{Conclusion}
In this paper, we strive to propose a simple and efficient network for lesion segmentation. Based on human process of carefully analyzed interference elimination, we propose a novel attention module, the DSA module, which consists of the FPSA branch and the FNSA branch, and utilize it to establish a pyramid CNN decoder with simple structure. Employing this decoder and the pre-trained MiT backbone encoders, we implement four scales of DSNet.  Extensive quantitative results on three public datasets demonstrate the effectiveness of DSNet, which achieves a trade-off between accuracy and computational complexity. We also conduct ablation studies to  further illustrate the validity of DSA module. In the future, we will conduct further experiments with DSNet to explore its application in real-time segmentation of various lesions and optimise its architecture to improve its performance. We hope that DSNet can be widely applied in the task of lesion segmentation, and moreover, our design can bring more inspiration to other researchers.
\vfill\pagebreak
\bibliographystyle{paper}
\bibliography{refs}
\end{document}